# On the characteristic length scale for the synthetic turbulence based on the Spalart-Allmaras model

Qilong Guo[1], Pengxin Liu[1], Chen Li[1] *, Dong Sun[1], Xianxu Yuan[1]

*1. State Key Laboratory of Aerodynamics, Mianyang 621000, People's Republic of China*

**ABSTRACT**: In the hybrid RANS-LES simulations, proper turbulent fluctuations should be added at the RANS-to-LES interface to drive the numerical solution restoring to a physically resolved turbulence as rapidly as possible. Such turbulence generation methods mostly need to know the distribution of the characteristic length scale of the background RANS model, which is important for the recovery process. The approximation of the length scale for the Spalart-Allmaras (S-A) model is not a trivial issue since the model's one-equation nature. As a direct analogy, the approximations could be obtained from the definition of the Prandtl's mixing length. Moreover, this paper proposes a new algebraic expression to approximate the intrinsic length scale of the S-A model. The underlying transportation mechanism of S-A model are largely exploited in the derivation of this new expression. The new proposed expression is employed in the generation of synthetic turbulence to perform the hybrid RANS-LES simulation of canonical wall-bounded turbulent flows. The comparisons demonstrated the feasibility and improved performance of the new length scale on generating synthetic turbulence at the LES inlet.

## 1. Introduction

Hybrid Reynolds Averaged Navier-Stokes (RANS) - Large Eddy Simulation (LES) methods have been kept improving during the last decades. They have combined the high-efficiency of RANS method and the capability of LES method to resolve large scale turbulent structures. In the hybrid methods, LES is only employed in the region where the large scales need to be resolved, and RANS method is used to model the mean flow in the rest regions. There are many strategies to operate the hybridization in the literature. Frohlich and von Terzi [1] summarized the basic concepts, the classification and the limitations of the hybrid RANS-LES methods. When the hybrid methods are applied in a zonal/embedded way, there always exist some artificial interfaces between the regions of both methods. The coupling boundary conditions on such interfaces are important since the performance of the LES on the downstream of the interface would greatly depend on the features of upstream unsteady flows. Proper turbulent fluctuations should be added at the interface, otherwise there would exist a large adaptation region contained in the LES region for building a physically resolved turbulence. An oversized adaptation region can degrade the accuracy of the solution in the whole downstream LES region.

---

*Corresponding authors: lichen@skla.cardc.cn

The RANS methods based on the statistical averaging of the Navier-Stokes (N-S) equations only solve the mean flow and compute the influence of the turbulence statistics by semi-empirical models. The approximated mean flow and low-order statistics can be used to synthesize the turbulent fluctuations for the LES inlet. A common fundamental principle for synthesizing the fluctuations is that the statistical information based on the RANS results must approximate the real physical turbulence as close as possible. The statistical information given by RANS can be employed through various methods such as the synthetic turbulence generator (STG) [2, 3], the synthetic eddy method (SEM) [4, 5], the synthetic Fourier modes methods [6, 7] and the dynamic forcing method [8]. Comprehensive reviews have been given by Tabor and Baba-Ahmadi [9], Dhamankar et al. [10], and Wu [11].

Recently, Probst et al. [12] evaluated the performances of SEM and STG as the grey area mitigation tools in the wall-bounded turbulent flow with mild separation. It is shown that such synthetic fluctuations are indeed helpful for improving the accuracy of the hybrid RANS-LES computation. Patterson et al. [13] studied the bias and temporal convergence errors of STG when used to generate the inflow of direct numerical simulation (DNS). An explicit method to measure these errors introduced by the random number arrays is developed, which can be employed to obtain an optimized selection of the random numbers with minimized errors. Generally, the basic input for this kind of methods are the limited statistical information obtained from the RANS computation. To this end, the two-equation RANS models are naturally superior to the one-equation RANS models since they contain the modeling for the independent transportations of two characteristic scales, which allows approximating the second-order statistics without any ambiguity. But for the one-equation RANS models, only one single transport equation is directly solved for representing one characteristic scale. Thus, the proper approximation of the second-order statistics relies on supplementing proper algebraic expression for the other characteristic scale (usually the characteristic length scale). Hence, it is not straightforward to use the one-equation models in cooperation with the SEM or the STG. Indeed, we can find that the background RANS models in the literatures are mostly the two-equation models [2-5].

In the community of aerospace engineering, the one-equation model proposed by Spalart and Allmaras [14], referred as S-A model hereafter, has been one of the most successful turbulence models in the last decades. Therefore, there exists a strong potential of the S-A model for playing the role of the RANS part in the hybrid simulations. This motivates the present work on surveying the approximation of the length scale for the S-A model, which is indispensable for the employment of the LES inlet treatment like the aforementioned SEM or STG.

In this paper, we choose the STG in [2, 3] to generate the artificial turbulent fluctuations imposed on the LES inlet. The STG is based on a superposition of a series of randomly generated Fourier modes with all the random quantities defined only once at the beginning of the simulation, which makes the method simple and efficient to implement. The second-order statistics needed by the STG is a prescribed model energy spectrum, which is crucial for constructing the Fourier modes. The model energy spectrum is characterized by a single length scale representing the wave length of the most energy-containing mode. As mentioned above, this length scale can be easily obtained for the two-equation model such as the $k$-$\omega$ Shear Stress Transport model [15], referred as SST model hereafter. When the background RANS is changed to S-A model, an algebraic expression should be found to approximate this length scale. Even if a feasible algebraic expression is provided, it cannot provide sufficient consideration about the influences from the historical and boundary information without an explicit modeling for the transportation equation related to this length scale. This flaw brought by the algebraic expression might affect the recovery process of the synthetic turbulence to the realistic one.

Concerning the above issues, the objective of this paper is to provide reasonable approximations of the length scale as the input of the STG for the S-A model, and to investigate the connections between the length scale and the recovery process. It will be shown that feasible approximations can be obtained from the definition of the Prandtl's mixing length. As an alternative way, it is argued that an intrinsic equation for the length scale of S-A model can be derived from the transportation equation. The numerical tests on the canonical wall-bounded turbulent flows suggest that the recovery process can be improved through employing the new proposed expression.

The paper is organized as follows. The main procedures of the STG and the role played by the length scales are introduced in Section 2. In Section 3, we present some feasible approximations of the length scale for S-A model as well as the derivations of the novel algebraic expression of the turbulent length scale. Moreover, a posteriori test is performed based on the highly-resolved simulation data of a turbulent boundary layer. In Section 4, the expressions of the length scale are adopted in the STG [2, 3] based on the S-A model to construct the inflow turbulent fluctuations for the hybrid RANS-LES simulation. Two canonical wall-bounded turbulent flow cases, i.e. the flat-plate boundary layers and fully-developed channel flows, are employed to make comparisons on the skin frictions, the second-order statistics and the spanwise integral length scale downstream of the interface. The conclusions are given in section 5.

## 2. Synthetic Turbulence Generation Method

2.1 *Basic formula*

The STG method in [2, 3] has been successfully applied in the simulation of several turbulence problems with the SST model. The main procedure is to construct the turbulent velocity fluctuations from the Reynolds stress tensor computed by the RANS model at the inlet as

$$u'_i(\vec{r}, t) = a_{ij} v'_j(\vec{r}, t), \tag{1}$$

where $u'_i$ for $i=1,2,3$ are corresponding to the component of the velocity fluctuation on $x_i$-directions. $a_{ij}$ is defined by the Cholesky decomposition of the Reynolds stress tensor $R_{ij}$, which is given as,

$$a_{ij} = \begin{bmatrix} \sqrt{R_{11}} & 0 & 0 \\ R_{21}/a_{11} & \sqrt{R_{22} - a_{21}^2} & 0 \\ R_{31}/a_{11} & (R_{32} - a_{21}a_{31})/a_{22} & \sqrt{R_{33} - a_{31}^2 - a_{32}^2} \end{bmatrix}. \tag{2}$$

The deviatoric stresses are computed by the Bousinessq approximation, while the normal stresses equal to 2/3 of the turbulent kinetic energy $k$. The vector $v'_j$ is constructed by superposing the spatiotemporal Fourier modes weighted by the normalized amplitudes on a von Karman spectrum model $E(\xi)$, which are given as

$$\mathbf{v}'(\vec{r}, t) = \sqrt{6} \sum_{n=1}^{N} \sqrt{q^n} \left[ \boldsymbol{\sigma}^n \cos(\xi^n \mathbf{d}^n \cdot \mathbf{r}'(t) + \varphi^n) \right], \tag{3}$$

$$q^n = \frac{E(\xi^n)\Delta\xi^n}{\sum_{n=1}^{N} E(\xi^n)\Delta\xi^n}. \tag{4}$$

where the superscript '$n$' stands for the $n$-th Fourier modes, $q$ is the local amplitude, and $\boldsymbol{\xi} = \xi \cdot \mathbf{d}$ is the wavenumber vector with the magnitude being $\xi$. $\mathbf{d}$ is a random unit vector of direction uniformly distributed over a sphere. $\boldsymbol{\sigma}$ is another random vector perpendicular to $\mathbf{d}$. $\varphi$ is the random phase uniformly distributed in $[0, 2\pi]$. The time-dependent vector $\mathbf{r}'(t)$ and the model spectrum $E(\xi)$ are given by

$$\mathbf{r}' = \{x'_1, x'_2, x'_3\}, \quad x'_1 = \frac{2\pi}{\xi^n \cdot \max\{L_e(r)\}} (x_1 - U_0 t), \quad x'_2 = x_2, \quad x'_3 = x_3, \tag{5}$$

$$E(\xi) = \frac{(\tilde{\xi}_e)^4}{\left[1 + 2.4(\tilde{\xi}_e)^2\right]^{17/6}} \cdot f_\eta \cdot f_{cut} \quad \text{with } \tilde{\xi}_e \equiv \xi \cdot L_e / 2\pi, \tag{6}$$

where $U_0$ is the macro-scale velocity at the interface, $f_\eta$ and $f_{cut}$ are empirical functions and $L_e$ is the length scale of the most energy-containing mode. Other detail expressions for the above quantities and functions can be referred to the reference [3].

In the above equations, the quantities provided by the background RANS model are the turbulent kinetic energy $k$, the turbulent eddy viscosity $\nu_t$ and the characteristic length scale $L_e$. The former two quantities ($k$ and $\nu_t$) are used for forming the matrix $a_{ij}$ (or $R_{ij}$), while the characteristic length scale $L_e$ is used for the description of the normalized model energy spectrum in Eq. (6), which is discussed in the next subsection.

2.2 *The model energy spectrum*

The determination of the model energy spectrum requires that at least two characteristic scales, i.e. one characterizing the kinetic energy and the other one providing length scale, must be known from the RANS computation.

The scale related to the turbulent kinetic energy is definite and straightforward for the RANS models based on the eddy-viscosity hypothesis since the transportation equation of the turbulent kinetic energy can be derived from the N-S equation with relatively less empiricism. In fact, the transportation of the kinetic energy is directly modeled in some RANS methods, such as $k$-$\varepsilon$ model, $k$-$\omega$ model, etc. For the models which do not directly involve the turbulent kinetic energy, it can be obtained from the assumption that the ratio of the Reynolds shear stress to the turbulent kinetic energy is a constant [16], i.e. $R_{12} \approx \beta_r k$, where $\beta_r = 0.3$ is the Bradshaw's constant. Considering the Boussinesq's hypothesis, the turbulent kinetic energy can be approximated as

$$k \approx \frac{2\nu_t S_{12}}{\beta_r} \approx \frac{\nu_t S}{\beta_r}, \tag{7}$$

where $S = \sqrt{2S_{ij}S_{ij}}$, $S_{ij} = 0.5(\partial u_i/\partial x_j + \partial u_j/\partial x_i)$. The profiles of $k$ for a turbulent boundary layer with $Re_\theta = 900$ are examined in Fig. 1. The results obtained from the S-A model are compared to the $k$-profile from the SST model. The profile of $R_{12}/\beta_r$ from the direct numerical simulation (DNS) data of Wu and Moin [17] is also shown for comparison. It can be seen that the approximation for the turbulent kinetic energy of the S-A model is acceptable.

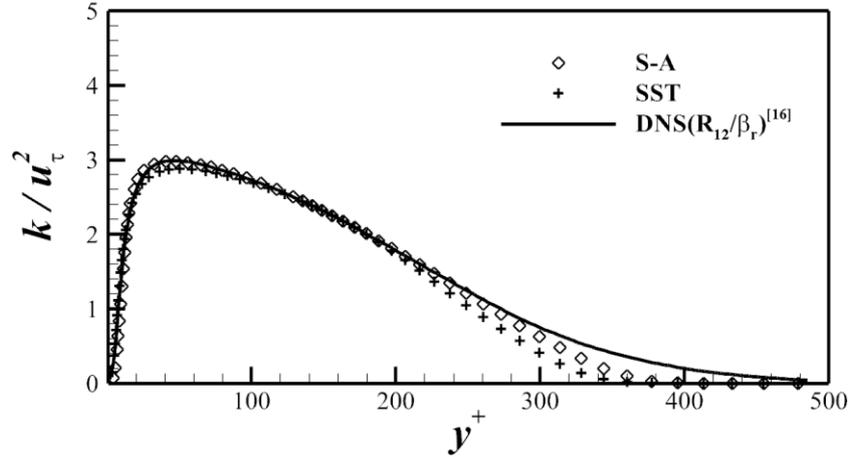

Fig. 1 The distributions of the turbulent kinetic energy for S-A and SST models in a turbulent boundary layer

The other characteristic scale represents the wavelength of the most energy-containing mode, i.e. the $L_e$ in Eq. (6). If adequate information, either by scale resolving simulation or by experiment measurements, about the turbulent statistics are provided, multiple integral quantities characterizing the length scales in turbulence can be defined. Following [18, 19], the integral length scale obtained from a real von Karman model spectrum is almost equivalent to the one corresponding to the most energy-containing mode in the most idealized situation, i.e. the homogeneous isotropic turbulence. When the wall-bounded turbulence is concerned, these length scales are functions of the spatial coordinates. For the SST model, it is suggested in [2] that $L_e$ can be approximated as

$$L_e = min(2d_w, C_l \cdot L_t), \tag{8}$$

where $d_w$ is the minimum distance to the wall, $L_t$ is some well-defined functional length scale (also called as the turbulent length scale) based on the known variables from the RANS computation, and $C_l$ =3.0. The term $2d_w$ agrees with Townsend's model [20] which states that the length scale of the large eddy in a turbulent boundary layer is proportional to the wall-normal distance. When the background RANS for the STG is the SST model, the turbulent length scale can be easily obtained from the known quantities ($k$ and $\omega$),

$$L_t = k^{1/2}/(C_\mu \omega), \tag{9}$$

with $C_\mu$=0.09. However, when the one-equation S-A model is employed as the background RANS model, the issue about approximating the turbulent length scale is not as straightforward as for the two-equation models as discussed in section 1. If we want to successfully apply the STG method in conjunction with the S-A model, the turbulent length scale must be properly approximated as an algebraic expression of the known quantities.

## 3. Length scales for the S-A model

In this section, we discuss some available choice for approximating the length scale in order to realize the coupling between the S-A model and the STG. The discussion is limited in the category of the wall-bounded turbulent flows since their intensive engineering interests. Besides the above closure issue, it is also natural to ask what behavior should the input $L_e$ has along the boundary layer to achieve the best recovery process for the STG. Moreover, we also propose a new algebraic expression for the turbulent length scale from the transport equation of S-A model.

*3.1 algebraic expression for the length scale*

(i) analogy to the Prandtl's mixing length

In a turbulent boundary layer, the Prandtl's mixing length $l_m$ is defined as

$$\nu_t = l_m^2 \cdot S. \tag{10}$$

As a general available approximation for the characteristic length scale for the RANS model, an analogy between the turbulent length scale and $l_m$ can be reasonably assumed, which gives

$$L_t \propto l_m = \sqrt{\frac{\nu_t}{S}}. \tag{11}$$

The effectiveness of the Prandtl's mixing length is partly confirmed by the success of some zero-equation models. For a turbulent boundary layer, experimental measurements have shown that $l_m = \kappa d_w$ holds in the log-layer, and $\kappa=0.41$ is the von Karman constant. Therefore, we can rewrite it as

$$L_t = \frac{2}{\kappa} l_m = \frac{2}{\kappa} \sqrt{\frac{v_t}{S}}. \tag{12}$$

in order to fitting Townsend's attached eddy model ($L_t \approx 2d_w$) in the near-wall region.

(ii) modified Prandtl's mixing length

According to a posteriori test about the length scale in the next subsection, a constant coefficient is employed to make a correction on Eq. (12), which becomes

$$L_t = \frac{2.7}{\kappa} \sqrt{\frac{v_t}{S}}. \tag{13}$$

(iii) Townsend's attached eddy model

Following the original recommendation in [2, 3] (also see Eq. (8)), the length scale is further limited as not larger than the Townsend's attached eddy model, i.e. $L \approx 2d_w$. Thus, we have the following equation:

$$L_e = min\left(2d_w, C_l \cdot \frac{2}{\kappa} \sqrt{\frac{v_t}{S}}\right), \tag{14}$$

with $C_l = 3.0$.

(iv) new expression

Since the primary mechanism of the transportation of the turbulent viscosity has been well modeled and carefully calibrated by the model equation, it is reasonable to relate the length scale to the transport equation. Based on this idea, we proposed a new algebraic expression for the turbulent length scale. The derivation is given in the following.

First, the dynamical behavior of the turbulent kinetic energy is examined by written its transport equation as

$$\frac{Dk}{Dt} = \tau_{ij} \frac{\partial u_i}{\partial x_j} - \varepsilon + \frac{\partial}{\partial x_j}\left[v \frac{\partial k}{\partial x_j} - \frac{1}{2}\overline{u_i' u_i' u_j'} - \frac{1}{\rho}\overline{p' u_j'}\right], \tag{15}$$

where $\frac{D(\cdot)}{Dt} \equiv \frac{\partial(\cdot)}{\partial t} + u_j \frac{\partial(\cdot)}{\partial x_j}$, and $\varepsilon$ is the turbulence dissipation rate. It should be noticed that Eq. (15) is derived from the transport equation of Reynolds-stress tensor without introducing any approximation. It can be also regarded as the theoretical basis for many eddy-viscosity models. With the Reynolds' average operation, the dissipation of the kinetic energy in the turbulence cascading is modeled by the single term $\varepsilon$. Therefore, the description for the dissipation of $k$ can be assumed as $Dk/Dt = -\varepsilon$, which indicates that the time scale for the vanishing of the turbulent kinetic energy can be characterized by $\tilde{t} = k/\varepsilon$. Thus, the averaged spatially travelling length for an energy-containing eddy before dissipated out is

$$L = \tilde{u} \cdot \tilde{t} = k^{3/2}/\varepsilon, \tag{16}$$

where $\tilde{u} = \sqrt{k}$ is the characteristic velocity. It has been shown in [7] that the length scale(in definition of $k^{3/2}/\varepsilon$) can be approximated by $2d_w$ in the near wall region. Indeed, Eq. (16) gives the proper estimation of the turbulent length scale for the standard $k$-$\varepsilon$ model [21].

Based on the above understanding, we reinterpret the turbulent length scale of S-A model as the same expression of the modeled kinetic energy and the turbulence dissipation rate, i.e. Eq. (16). If the turbulence dissipation rate $\varepsilon$ is appropriately provided, the turbulent length scale is determined with the turbulent kinetic energy given by Eq. (7). For example, if we define $\varepsilon_0$ as

$$\varepsilon_0 = C_\varepsilon \nu_t S^2, \tag{17}$$

with $C_\varepsilon = \beta_r^{-3/2} \cdot \kappa/2 \approx 1.248$, it is easy to find that Eq. (12) can be immediately obtained through substituting Eq. (17) into Eq. (16).

In order to seek proper expression for the dissipation rate for the S-A model, the model transport equation is also examined, which is given as

$$\frac{\partial \tilde{\nu}}{\partial t} + u_j \frac{\partial \tilde{\nu}}{\partial x_j} = \underbrace{C_{b1}\tilde{S}\tilde{\nu}}_{production} - \underbrace{C_{w1}f_w(\tilde{\nu}/d_w)^2}_{destruction} + \frac{1}{\sigma}\left[\frac{\partial}{\partial x_j}\left((\nu+\tilde{\nu})\frac{\partial \tilde{\nu}}{\partial x_j}\right) + C_{b2}\frac{\partial^2 \tilde{\nu}}{\partial x_j^2}\right], \tag{18}$$

where $\tilde{\nu}$ is the modeled transported variable, $\nu$ is the molecular viscosity, and $\tilde{S}$ is the modified strain rate defined by

$$\tilde{S} = S + \frac{\tilde{\nu}f_{v2}}{\kappa^2 d_w^2}. \tag{19}$$

The turbulent eddy viscosity $\nu_t$ is given by

$$\nu_t = \tilde{\nu}f_{v1}. \tag{20}$$

Functions $f_{v1}$, $f_{v2}$ and $f_w$ are empirical correlations and $C_{b1}$, $C_{b2}$, $C_{w1}$ and $\sigma$ are empirical parameters. The specific expressions and values of these quantities can be referred to [14].

By combining Eq. (7) and Eq. (20), we can obtain

$$\frac{Dk}{Dt} \cong \frac{D}{Dt}\left(\frac{Sf_{v1}}{\beta_r}\tilde{\nu}\right). \tag{21}$$

The empirical correlation $f_{v1}$ is the ratio of $\tilde{\nu}$ to $\nu_t$ and just equals 1.0 in the most part of the boundary layer(except in the viscous sub-layer and near the outer edge). Thus, Eq. (21) can be further reduced to

$$\frac{Dk}{Dt} \cong \frac{1}{\beta_r}\left(S\frac{D\tilde{\nu}}{Dt} + \nu_t\frac{DS}{Dt}\right). \tag{22}$$

The terms $\frac{Dk}{Dt}$ and $\frac{D\tilde{\nu}}{Dt}$ represent the convection effects of Eq. (15) and Eq. (18) respectively. The second term on the right hand $\nu_t\frac{DS}{Dt}$ corresponds to the diffusion brought by the turbulent eddy viscosity. To build the connection between Eq. (15) and Eq. (18), it is argued that the production and destruction terms contained in the left and right hand in Eq. (22) should obey the same behavior, which gives

$$\nu_t S^2 - \varepsilon \cong \frac{S}{\beta_r}[C_{b1}\tilde{S}\tilde{\nu} - C_{w1}f_w(\tilde{\nu}/d_w)^2] \cong \frac{S}{\beta_r}[C_{b1}S\nu_t - C_{w1}f_w(\nu_t/d_w)^2], \tag{23}$$

where the simplification $\tau_{ij}\frac{\partial u_i}{\partial x_j} = \nu_t S^2$ is employed. Then the approximation for the dissipation rate can be obtained as:

$$\varepsilon_1 = \left(1 - \frac{C_{b1}}{\beta_r}\right)\nu_t S^2 + \frac{C_{w1}}{\beta_r}f_w S(\nu_t/d_w)^2. \tag{24}$$

It should be emphasized that Eq. (24) is only reasonable in the regions where the approximation $f_{v1} \cong 1.0$ ($\tilde{\nu} = \nu_t$) holds. The empirical function $f_w$ can be further replaced by a constant 0.5 to simplify the calculation, which gives a rough approximation for this region. In the near-wall region, a hybrid operation with $\varepsilon_0$ is employed, namely

$$\varepsilon_h = (\varepsilon_0)^{p_w} \cdot (\varepsilon_1)^{1-p_w}, \tag{25}$$

$$p_w = 1.0 - \tanh[0.04 \cdot (1 + A \cdot \chi_t^2)], \tag{26}$$

where $\chi_t = \nu_t/\nu$ and $A = 1.3 + 3.8\frac{\nu}{S \cdot d_w^2}$. The power function $p_w$ is close to 1.0 in the near-wall region. Thus the hybrid dissipation $\varepsilon_h$ would return to $\varepsilon_0$ (Eq. (17)). In the outer layer of a turbulent boundary layer, $\varepsilon_h$ is equivalent to $\varepsilon_1$ (Eq. (24)). Fig. 2 presents the distribution of $\varepsilon_0$, $\varepsilon_1$ and $\varepsilon_h$ along a turbulent boundary layer profile with $Re_\theta = 900$. The power function $p_w$ is also presented.

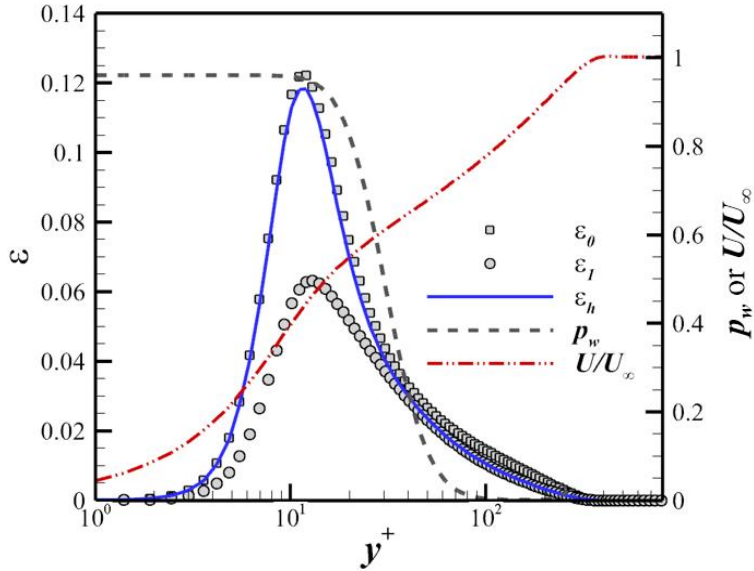

Fig. 2  Distributions of the approximated dissipation rates and the power function in a turbulent boundary layer

By substituting $\varepsilon_h$ into Eq. (16), a new algebraic expression for the length scale can be obtained. In order to avoid division by zero in Eq. (16), the minimum value of $\varepsilon_h$ should be constrained. Therefore, we can obtain the following expressions:

$$\varepsilon_h = \max\{(\varepsilon_0)^{p_w} \cdot (\varepsilon_1)^{1-p_w}, 1.0 \times 10^{-10}\}, \tag{27}$$

$$L = k^{3/2}/\varepsilon_h. \tag{28}$$

To make a summary, $\varepsilon_0$ in Eq. (17) comes from the length scale Eq.(12), which adopts the analogy expression of the Prandtl's mixing length as well as the Townsend's model. $\varepsilon_1$ in Eq. (24) is derived from the dissipative behavior contained in the transportation equation of the S-A model. And $\varepsilon_h$ in Eq. (28) represents a zonal hybrid between the two former expressions. It is worth noting that there are no non-local variables involved in the calculation of the above expressions (Eq. (12-14) and Eq. (28)).

*3.2 A posteriori test*

A zero pressure gradient turbulent boundary layer (abbreviated as TBL) over a flat plate is utilized here to perform a posteriori test about the distributions of the length scale from different algebraic expressions. The streamwise location of *Re$_\theta$*=1000 in the TBL is selected to perform the test.

Using the expressions (Eq. (12-14) and Eq. (28)) introduced in section 3.1, the distributions of the length scale can be extracted without any ambiguity. The background RANS data is obtained by the one-equation S-A model. In Fig.3, the grey-colored scatters are the data from several streamwise locations in a highly-resolved LES, which utilized very fine mesh approaching to the level of DNS (also called as quasi-DNS [22]). The results of the length scales are computed as [7]

$$L = \frac{2\sqrt{\langle u_1' u_2' \rangle}}{\kappa \cdot |du_1/dx_2|}. \tag{29}$$

The angular bracket stands for a spanwise and temporal average. The selected locations can be viewed as 'recovered states' compared to the STG inlet through justifying the skin friction and the profiles of second-order statistics. The details of the simulation will be elucidated in section 4. The plus symbols '+' in Fig. 3 are the results summarized by Glegg et al. (see Fig. 2 in [7]), which appear good agreement with the highly-resolved LES data. Hence, we refer to the data in [7] and in present computation as 'reference data' hereafter. For each data set, the length scale (*L*) and the distance to the wall (*x$_2$*) are both scaled by the local boundary layer thickness *δ*.

In Fig. 3, it is shown that significant discrepancies exist among the distributions of the approximated expressions. The first one, i.e. the analogy to the Prandtl's mixing length (Eq. (12)), is lower than the reference data in the most part of the boundary layer thickness. Besides, we could easily obtain Eq. (13) by adjusting the value of the coefficient to have the better fit to the reference data. The largest distribution in Fig. 3 is given by Eq. (14), which is directly inherited from the recommendation in [2, 3]. The curve coincide with the straight line of $L = 2d_w$ over a wide range of the boundary layer. It is also noted that reducing the value of $C_l$ to 1.35 would result in the same curve as Eq.(13). As a new proposal, Eq. (28) tries to build a reasonable relation between the length scale and the transportation behavior underlying the model equation. It agrees well with the reference data in the region of $x_2/\delta \leq 0.35$, but appears larger than the reference data in the outer part of the BL. Given the distribution of the turbulent kinetic energy, the wavelengths of the synthesized Fourier Modes are closely related to the adopted length scale expressions. In the next section, the four algebraic expressions for the length scale will be compared in the practical numerical experiments, which is aiming at shedding some lights on the question about which distribution is the most suitable one for the recovery of the inlet ST.

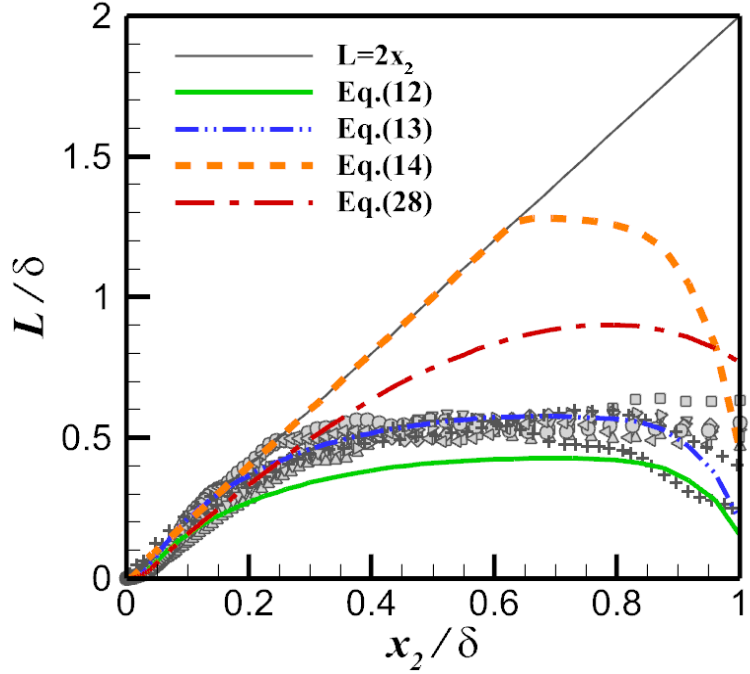

Fig. 3  Distributions of the length-scale expressions inside a turbulent boundary layer

(Grey-colored scatters: current computation; symbols '+': data in [7])

## 4. Numerical tests

In this section, two canonical wall-bounded turbulent flow cases are employed to perform the numerical tests. The first case is the TBL utilized in section 3.2, and the second one is a fully developed turbulent channel flow (abbreviated as TC). In the simulations, the (filtered) compressible N-S equations are solved using our in-house CFD solver, which has been validated in many complex flow cases [23-27]. The numerical methods employed in this paper are as following: the convection term is discretized by an optimized fourth-order finite-difference scheme which is duplicated from the linear form of WENO-SYMBO scheme [28]. The viscous term is discretized by the fourth-order central difference scheme. The dual time stepping is employed for time marching. The Mach number is set as 0.2 to approximate the incompressible regime. The Improved Delayed Detached Eddy Simulation (IDDES) [29] based on the S-A model is employed as the sub-grid scale model. Both of the cases are treated as spatial-developing. The prescribed velocity fluctuations (generated by the STG method) are superimposed on the corresponding RANS solutions at the inlet. The walls in the two cases are set as no-slip and isothermal. The wall temperatures are 300K. Sponge layers are utilized in both cases at the downstream of the effective computational domain to damp the spurious reflections. The spanwise boundaries are treated as periodic boundaries. The comparisons are carried out between the inlet turbulent fluctuations installed with the four algebraic expressions, which are termed with 'L1-L4' (see Table 1).

**TABLE 1  Approximations for the length scale**

| Case name | Length scale |
|---|---|
| L1 | Eq. (12) |
| L2 | Eq. (13) |
| L3 | Eq. (14) |
| L4 | Eq. (28) |

*4.1 turbulent boundary layer*

In the TBL case, the computational domain is illustrated in Fig. 4, $\delta_0$ is denoted as the nominal thickness. The Reynolds number based on the momentum thickness $Re_\theta$ is about 1000 at the inlet. The size of the effective computational domain in the streamwise ($x_1$), wall-normal ($x_2$) and spanwise ($x_3$) directions are $26\delta_0$, $5.4\delta_0$ and $4.5\delta_0$ respectively.

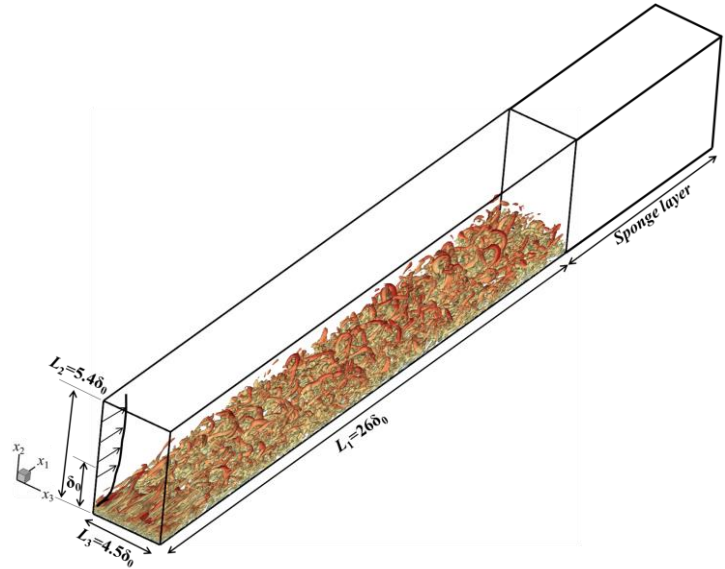

Fig. 4  Schematic of the computational domain of the TBL

In the current simulation, the grid spacings on the wall-parallel directions are approximately $\Delta x_1^+$=20.0 and $\Delta x_3^+$=10.0 in wall units, which is approaching DNS. The grid-point distributions on the wall-normal direction are kept same with the first spacing as $\Delta x_{2,min}^+$ =0.48. The number of grid points are $N_1 \times N_2 \times N_3$=521×121×181. In order to let the flow fields have sufficient time to evolve, all cases were run over approximately $150 \cdot \delta_0/U_\infty$ before the samples were collected, and the time-history of skin frictions was also examined to ensure the statistically stationary states had been reached.

The instantaneous fields of vorticity magnitude at the wall and $x_2^+$=15 are given in Fig. 5. The streamwise streaks attached to the wall can be clearly seen in the figure. No obvious qualitative difference is observed between the cases.

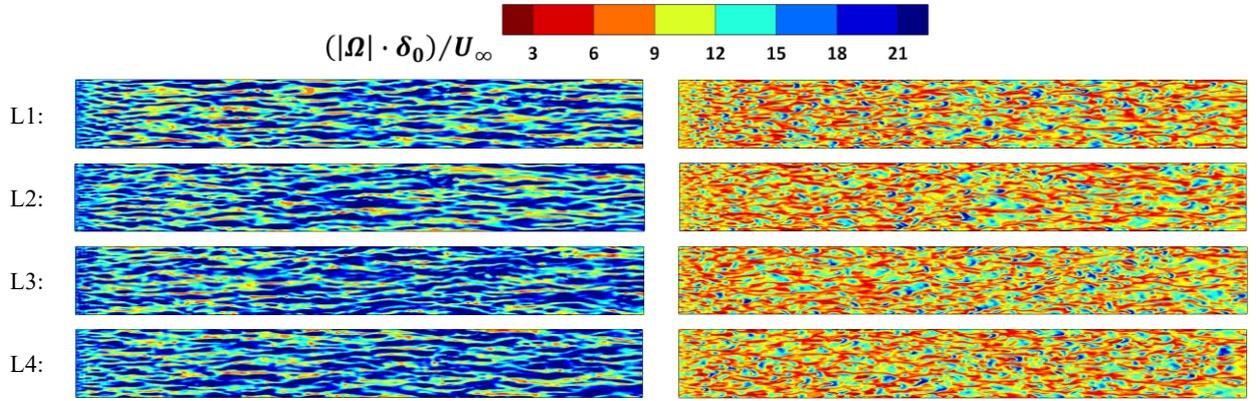

Fig. 5 Instantaneous fields of vorticity magnitude at $x_2$=0 (left panel) and $x_2^+$=15 (right panel)

In Fig. 6, the mean skin friction coefficients $C_f$ (versus both $Re_\theta$ and $x/\delta_0$) are compared with the empirical equation for the incompressible turbulent boundary layer, which is given as [30]

$$C_f = 0.024 \cdot Re_\theta^{-1/4}. \tag{30}$$

The symbols in the figure are the DNS data from [30, 31]. It can be seen that the shapes of $C_f$-curves displayed in both figures are similar. The skin friction coefficients in all the cases revert to their fully-developed level after a streamwise adaptation region of about $\Delta Re_\theta \approx$120~140 (or $\Delta x_1 \approx$8.4~9.6$\delta_0$) distance from the inlet. In the adaptation region, the value of $C_f$ is apparently lower than the reference value. Among the considered cases, the curves obtained from 'L3' and 'L4' show the best agreement with the reference value in the fully-recovered regions. The employment of 'L1' and 'L2' appear to decrease the $C_f$, whereas 'L2' is relatively closer to the reference value.

In order to examine the development of the quantitative characteristics of the TBL, we make a comparison on the mean velocity profiles and the second order statistics $R_{11}^+$, $R_{22}^+$ and $R_{12}^+$ ($R_{ij}^+ =$ $\langle u_i' u_j' \rangle / u_\tau^2$) at four streamwise locations, i.e. $x_1/\delta_0$=2.1, 4.2, 6.3 and 8.4. The corresponding $Re_\theta$ are about 1032, 1064, 1096 and 1128 respectively. The DNS data of Wu and Moin [17] are taken as the reference data. In Fig. 7, the mean velocity profiles at different streamwise locations in the TBL are presented. It can be seen that the discrepancies in the mean velocity profiles between L2, L3 and L4 are insignificant, while the profiles obtained from L1 are slightly higher than the other cases. This is caused by the underestimation of the $C_f$ in case L1 compared to L2, L3 and L4 (see Fig. 6). With the location moving downstream, apparent recovery processes can be observed. At the last location ($x_1/\delta_0$=8.4), the mean velocity profiles of L2, L3 and L4 are in very good agreements with the DNS data. It should also be mentioned that the differences of the values in the outer regions of the TBL compared to the DNS data are originated from the different $Re_\theta$ (or different $u_\tau$).

The Reynolds' stress components at the first position $x_1/\delta_0$=2.1 exhibit acceptable agreement with the reference data. By comparing different cases, it is found that the normal stress $R_{11}^+$ show substantial discrepancies, while the results of $R_{22}^+$ and $-R_{12}^+$ are slightly lower than the reference in all the cases. All the considered quantities grow larger when it moves to $x_1/\delta_0$=4.2. The results at $x_1/\delta_0$=6.3 and $x_1/\delta_0$=8.4 are very close to each other, indicating that the second-order statistics have already been fully-reverted. The most obvious differences between the cases are exhibited within wall-normal coordinate range 0.18<$x_2/\delta_0$<0.8 on the curves of $R_{11}^+$. The quantitate comparison suggest that the larger length scale will gives higher $R_{11}^+$. Recalling Fig. 3, this region is in accordance with the discrepant region of the provided length scales, indicating that the length scale for the inlet ST has an impact on the production of the turbulent kinetic energy.

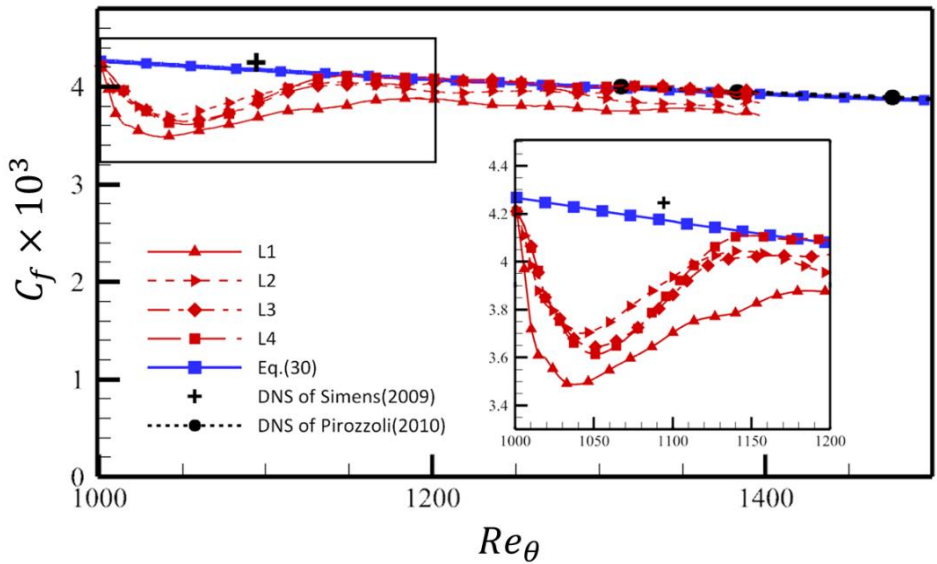

Fig. 6 Distributions of skin friction coefficients in TBL cases

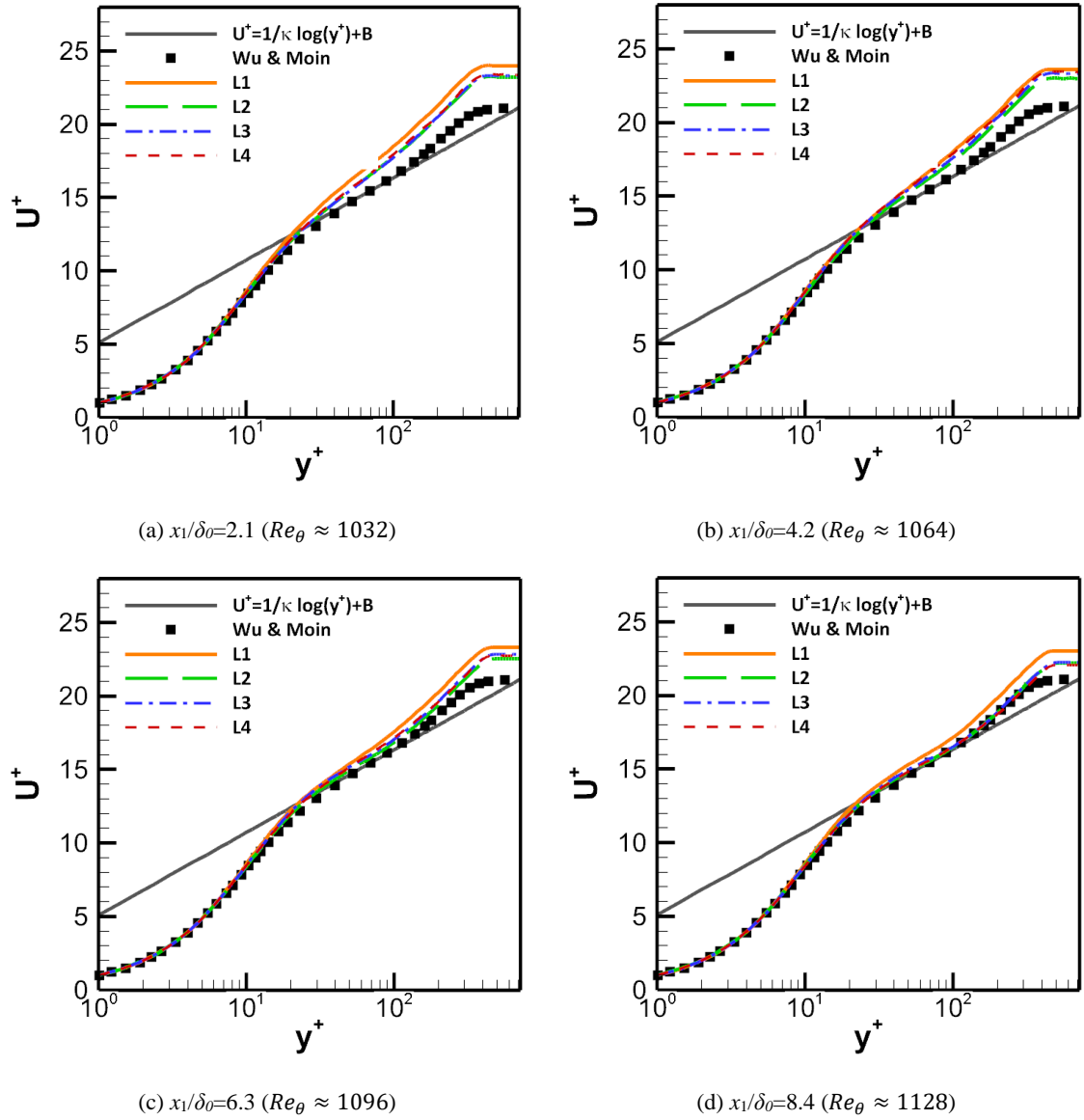

(a) $x_1/\delta_0$=2.1 ($Re_\theta \approx 1032$)  (b) $x_1/\delta_0$=4.2 ($Re_\theta \approx 1064$)

(c) $x_1/\delta_0$=6.3 ($Re_\theta \approx 1096$)  (d) $x_1/\delta_0$=8.4 ($Re_\theta \approx 1128$)

Fig. 7  Wall normal distributions of the mean velocity profiles at different steamwise locations in the TBL cases

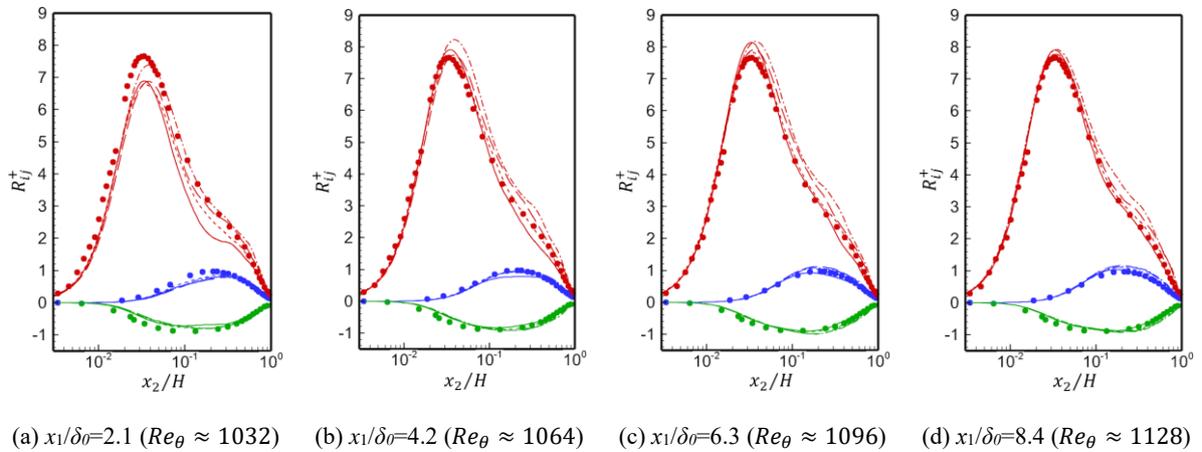

(a) $x_1/\delta_0$=2.1 ($Re_\theta \approx 1032$)  (b) $x_1/\delta_0$=4.2 ($Re_\theta \approx 1064$)  (c) $x_1/\delta_0$=6.3 ($Re_\theta \approx 1096$)  (d) $x_1/\delta_0$=8.4 ($Re_\theta \approx 1128$)

Fig. 8  Wall normal distributions of the Reynolds stresses at different streamwise locations in TBL cases

(Different cases: solid lines, L1; dashed lines, L2; dash-dot lines, L3; long dash lines, L4; solid circles, DNS [17].

Different colors: red, $R_{11}^+$; blue, $R_{22}^+$; green, $R_{12}^+$)

In order to survey the characteristic length scale of the vortices structures, the spanwise width length $l_z$ can be estimated as the distance between two successive crossing points on the two-point auto-correlation function with the constant 0.05 [32, 33], where the two-point auto-correlation function $\mathcal{R}$ is defined as

$$\mathcal{R}(x_1, x_2) = \frac{\langle u_1(x_1, x_2, x_3, t) \cdot u_1(x_1, x_2, x_3 + \Delta x_3, t) \rangle}{\langle u_1(x_1, x_2, x_3, t)^2 \rangle} \tag{31}$$

In Fig. 9, it is shown the distribution of $l_z$ along the wall normal direction at $x_1/\delta_0$=0.6 and $x_1/\delta_0$=11.2. The experimental data in [32] are also plotted. The curves from the cases indicate different characteristic length scales of the vortices structures even in the fully-recovered stage (symbols in Fig. 9, $x_1/\delta_0$=11.2) of the skin friction and the Reynolds' stresses. The result from case 'L4', among others, appears to give the closest distribution to the experimental data. At the near-inlet streamwise location (lines in Fig. 9, $x_1/\delta_0$=0.6), the estimated spanwise length scales $l_z$ are generally lower than the corresponding downstream results.

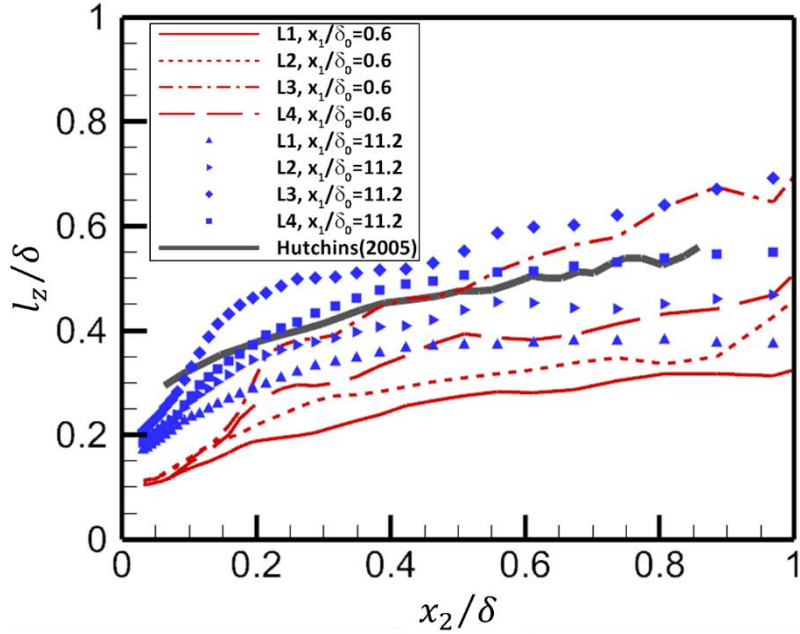

Fig. 9 Distribution of $l_z$ along the wall normal direction in TBL cases

*4.2 turbulent channel flow*

In the TC case, the Reynolds number based on the half-width ($H$) and the frictional velocity ($u_\tau$) at the inlet is about $Re_\tau$=395~400, and the numerical results of such case have been widely documented in literatures [34-36] with a periodic streamwise configuration. Here, the spatial-developing simulations are initialized using the mean-profile obtained by the streamwise periodic RANS simulation. A constant body force is added on the whole domain. The sizes of the computational domain are set as $L_{x1} \times L_{x2} \times L_{x3}$=28$H$×2$H$×3.4$H$. The adopted grid resolution in the wall-parallel directions are approximately $\Delta x_1^+$=28.0 and $\Delta x_3^+$=10.0 in wall units. In the wall-normal direction, the grids are clustered near the two walls and the minimum spacing in wall units is $\Delta x_{2,min}^+$=0.8. The number of points are $N_1 \times N_2 \times N_3$=401×109×137 in the streamwise, wall-normal, and spanwise directions respectively. The cases were run approximately $120 \cdot H/U_b$ before the samples were collected

There might exist a doubt about the lacking of necessity to perform numerical tests on the turbulent channel flows since the flow physics appear to be similar to a turbulent boundary layer. However, it is argued that the confined geometry configuration in the channel would lead to differences in the reaction to the inlet ST. When the flow is fully developed, the boundary layers attached on the two opposite walls would interact with each other and the influence of the inner layer (log-layer) becomes more critical.

The mean skin friction coefficients of the channel flow are compared with the RANS result ($C_{f,RANS} \approx$6.24×10$^{-3}$) in Fig. 10. The adaptation region lasts about 7~10$H$ in the streamwise direction. It is seen that the shortest recovery region is obtained by 'L4', while the case 'L2' is the latest one to reach the correct mean skin friction. The results show different trend with those in the turbulent boundary layer cases, indicating the existence of the discrepant reaction to the inlet ST as mentioned above. Four streamwise locations are chosen to make comparisons on the mean velocity profiles as well as the second order statistics $R_{11}^+$, $R_{22}^+$ and $R_{12}^+$ within $x_1/H$=2.5~10.9. The DNS data of Modesti et al. [34] ('INC6' case), which has the similar $Re_\tau$, are employed as reference. In Fig. 11, the mean velocity profiles obtained from different cases are compared. It is depicted that the L4 case achieves the earliest agreement with the DNS data with the location moving downstream. For the second order statistics, Fig. 12 clearly shows the streamwise variations in each component, especially near the peak of $R_{11}^+$. The best agreement with the DNS data is obtained by the result of 'L4'. The $R_{11}^+$ in case 'L2' is significantly larger than the DNS data, while the $R_{11}^+$ in cases 'L1' and 'L3' fall in between those of 'L2' and 'L4'.

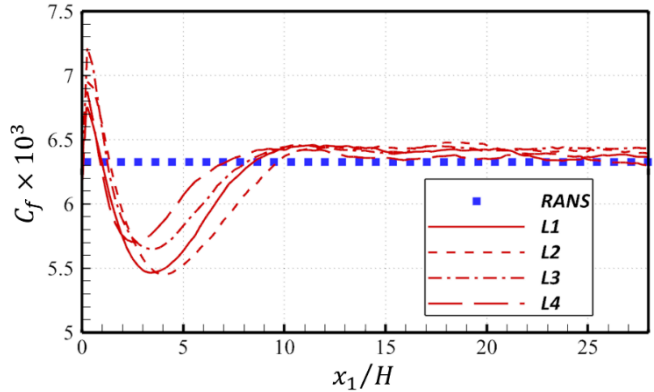

Fig. 10   Streamwise distributions of skin friction coefficients in the channel flow

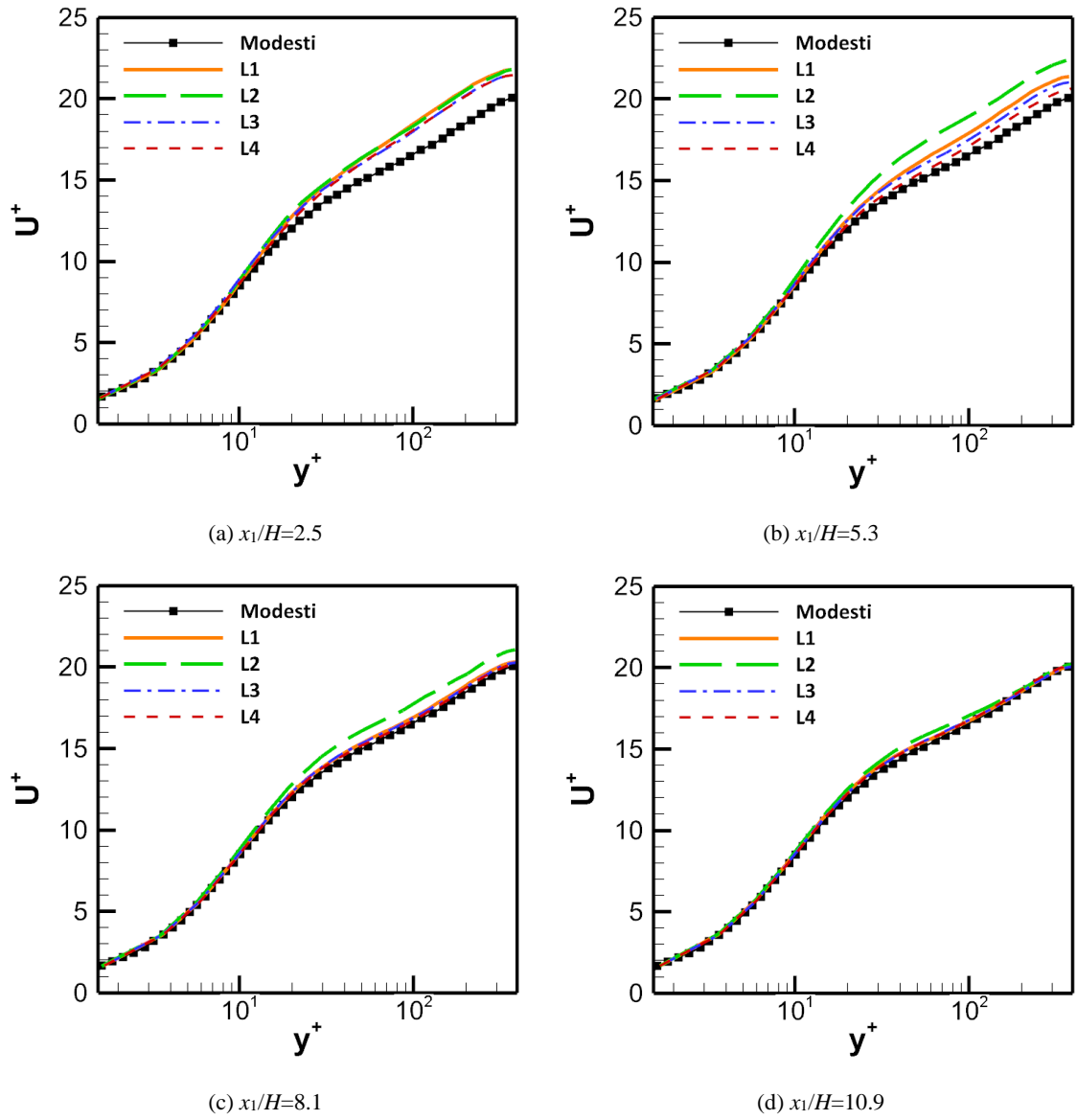

Fig. 11 Wall normal distributions of the mean velocity profiles at different steamwise locations in the TC cases

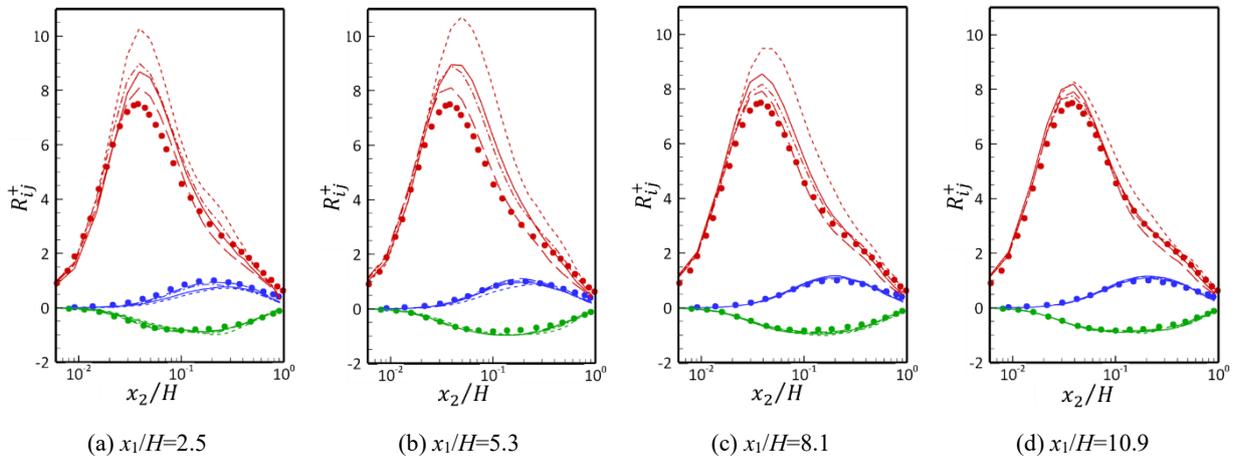

(a) $x_1/H$=2.5  (b) $x_1/H$=5.3  (c) $x_1/H$=8.1  (d) $x_1/H$=10.9

Fig. 12  Wall normal distributions of the Reynolds stresses at different streamwise locations in TC cases

( Different cases: solid lines, L1; dashed lines, L2; dash-dot lines, L3; long dash lines, L4;

solid circles, DNS [34]. Different color: red, $R_{11}^+$; blue, $R_{22}^+$; green, $R_{12}^+$ )

The distributions of the spanwise integral length scale $l_z$ along the two typical wall-normal lines ($x_1/H$=2.5 and 10.9) are given in Fig. 13. At the two streamwise locations, different relative trends between the cases are shown compared to the TBL: the results of 'L4' become close to those in 'L1' along the near-wall range of $x_1/H$=2.5 as well as the whole range of $x_1/H$=10.9. The experimental measurements from Monty et al. [33] are plotted in the grey solid line, and good agreements with the profiles from L1 and L4 can be observed.

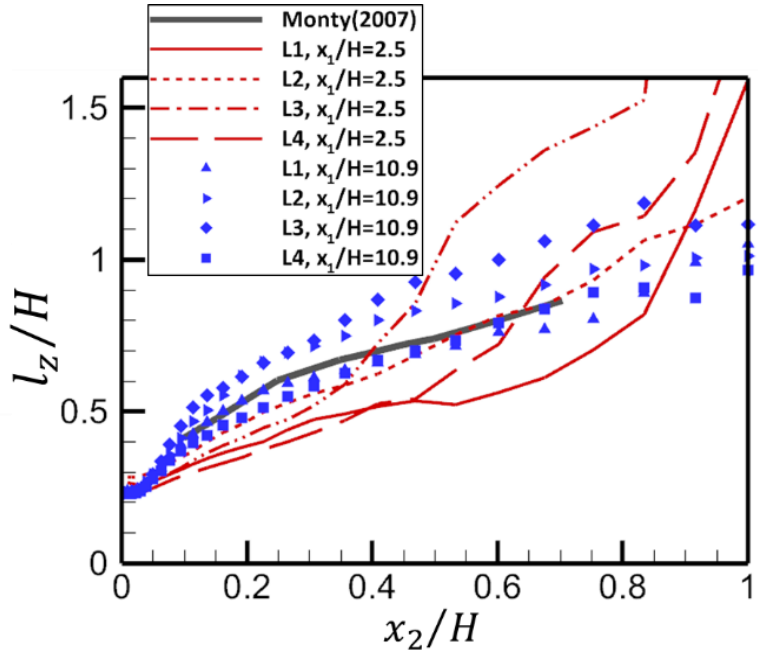

Fig. 13  Variation of the spanwise integral length scale with wall-normal coordinate

*4.3 length scale versus recovery process*

From the above analyses, it is demonstrated that the considered length scale expressions are feasible for providing the input information of the STG based on the one-equation model. The development of skin friction, second order statistics and the spanwise integral length scale in both the two wall-bounded flows are examined to investigate the recovery process.

Eq. (12) gives the longest recovery distance in TBL case, but its recovery distance in TC appears moderate. On the contrary, Eq. (13) results in the latest recovery in the TC computations, but its performance in TBL is good. This suggest that the Prandtl's mixing length is suitable for the near-wall region (log-layer). But in the outer layer, the Prandtl's mixing length may be relatively small in the outer layer. This is why a factor 1.35 can lead to the improved performance of Eq. (13) compared to Eq. (12). As the largest length scale, Eq. (14) gives satisfactory recovery in $C_f$ of both cases. Its main problems are the ill recovery process of $R_{11}^+$ in the TBL (see Fig. 8), and the larger spanwise length scale than the experimental measurements. Based on the above understanding, the new expression of Eq. (28) is equivalent to the Prandtl's mixing length in the inner layer, and properly exploits the dissipation mechanism contained in the S-A model. If considering all the comparisons, Eq. (28) gives the best recovery process among all the expressions.

## 5. Conclusion

The characteristic length scale plays an important role for generating proper turbulent fluctuations on the interface of the hybrid RANS-LES simulations. Based on the STG method [2, 3], we discussed the issue about the approximations of turbulent characteristic length scale for one-equation S-A model, which still remains unclosed. Unlike the two-equation eddy-viscosity models, the one-equation model cannot simultaneously describe both the turbulent kinetic energy and the turbulent characteristic length scale, the known of which are the most fundamental requirement for reproducing the reasonable turbulent statistics. Several available approximated expressions of the length scale, including a new expression representing the relation between the length scale and the underlying transportation mechanism of the model equation, are compared in a posteriori test using the highly-resolved LES data of a turbulent boundary layer.

A posteriori test shows that the analogy to the Prandtl's mixing length gives smaller distribution than the DNS data, while the original recommendation in references [2, 3] result in the largest distribution. Through adjusting the constant analogy coefficient, the agreement can be significantly improved. The new expression agrees well with the reference data in the region of $x_2/\delta \leq 0.35$, but becomes larger than the reference data in the outer part of the boundary layer. The performances of the inlet fluctuations generated by different expressions for turbulent length scale have been assessed on the turbulent boundary layer over a flat-plate and the spatially developed turbulent channel flow. The recovery processes of the skin friction, the second-order statistics, as well as the spanwise integral length scale are examined in detail. It is revealed that improved recovering behaviors have been achieved through adopting the new length scale expression for generating the inlet turbulent fluctuations compared to the other considered expressions.

## Acknowledgements

This work was supported by National Key Research and Development Program of China (2019YFA0405201), National Natural Science Foundation of China (12002360, 92052301) and National Numerical Windtunnel project.

# References

[1]  Frohlich, J., von Terzi, D.: Hybrid LES/RANS methods for the simulation of turbulent flows. *Progress in aerospace sciences*. **44**(5), 349-377 (2008)

[2]  Admian, D., Travin, A.: An efficient generator of synthetic turbulence at RANS-LES interface in embedded LES of wall-bounded and free shear flows. In: *Computational Fluid Dynamics 2010, Proceedings of the 6th international conference on computational fluid dynamics*, St. Petersburg, 739-744 (2010)

[3]  Shur, M., L., Spalart, P., R., Strelets, K., Travin, A.: Synthetic Turbulence Generators for RANS-LES interfaces in zonal simulations of aerodynamic and aeroacoustic problems. *Flow Turbulence and Combustion*. **93**(1), 63-92(2014)

[4]  Poletto, R., Craft, T., Revell, A.: A new divergence free synthetic eddy method for the reproduction of inlet flow conditions for LES. *Flow Turbulence and Combustion*. **91**(3), 519-539 (2013)

[5]  Jarrin, N., Prosser, R., Uribe, J., C., Benhamadouche, S., Laurence, D.: Reconstruction of turbulent fluctuations for hybrid RANS/LES simulations using a Synthetic-Eddy Method. *International Journal of Heat and Fluid Flow*. **30**(3), 435-442 (2009)

[6]  Auerswald, T., Probst, A., Bange, J.: An anisotropic synthetic turbulence method for large-eddy simulation. *International journal of Heat and Fluid Flow*. **62**(B), 407-422 (2016)

[7]  Glegg, S., Morin, B., Atassi, O., Reba, R.: Using Reynolds-Averaged Navier-Stokes calculations to predict trailing-edge noise. *AIAA Journal*. **48**(7), 1290-1301 (2010)

[8]  Laraufie, R., Deck, S., Sagaut, P.: A dynamic forcing method for unsteady turbulent inflow conditions. *Journal of Computational Physics*. **230**, 8647-8663 (2011)

[9]  Tabor, G., R., Baba-Ahmadi, M., H.: Inlet conditions for large eddy simulation: A review. *Computers and Fluids*. **39**(4), 553-567 (2010)

[10] Dhamankar, N., S., Blaisdell, G., A., Lyrintzis, A., S.: An overview of turbulent inflow boundary conditions for large eddy simulations (Invited). AIAA paper 2015-3213 (2015)

[11] Wu, X.: Inflow turbulence generation methods. *Annual review of fluid mechanics*. **49**, 23-49 (2017)

[12] Probst, A., Schwamborn, D., Garbaruk, A., Guseva, E., Shur, M., Strelets, M., Travin, A.: Evaluation of grey area mitigation tools within zonal and non-zonal RANS-LES approaches in flows with pressure induced separation. *International Journal of Heat and Fluid Flow*. **68**, 237-247 (2017)

[13] Patterson, J., W., Balin, R., Jansen, K., E.: Assessing and improving the accuracy of synthetic turbulence generation. *Journal of Fluid Mechanics*. 906, R1. (2021)

[14] Spalart, P., R., Allmaras, S., R.: A one-equation turbulence model for aerodynamic flows. AIAA paper 92-0439 (1992)

[15] Menter, F., R.: Zonal two-equation k-ω turbulence models for aerodynamics flows. AIAA paper 1993-2906 (1993)

[16] Wilcox, D., C.: Turbulenced modeling for CFD. 3rd ed., DCW Industries, 2006, Chapter 4

[17] Wu, X., Moin, P.: Direct numerical simulation of turbulence in a nominally zero-pressure-gradient flat-plate boundary layer. *Journal of Fluid Mechanics*. **630**, 5-41 (2009)

[18] Hunt, J., C., R.: A theory of turbulent flow round two-dimensional bluff bodies. *Journal of Fluid Mechanics*. **61**(4), 625-706 (1973)

[19] Carlotti, P., Drobinski, P.: Length scales in wall-bounded high-Reynolds-number turbulence. *Journal of Fluid Mechanics*. **516**, 239-264 (2004)

[20] Townsend, A., A.: The structure of turbulent shear flow, 2nd ed., Cambridge University Press, New York, 1976, Chapter 7

[21] Launder, B., E., Sharma, B., I.: Application of the energy dissipation model of turbulence to the calculation of flow near a spinning disc. *Letters in Heat and Mass Transfer*. **1**(2), 131-138 (1974)

[22] Spalart, P. R., Jou, W.-H., Strelets, M., Allmaras, S. R.: Comments on the feasibility of LES for wings, and on a hybrid RANS/LES approach. *Proceedings of 1st AFOSR International Conference on DNS/LES, Advances in DNS/LES*. 137-147 (1997)

[23] Liu, P.-X., Guo, Q.-L., Sun, D., Li, C., Zhang, H.-X.: Wall Effect on the Flow Structures of Three-dimensional Rotating Detonation Wave. *International Journal of Hydrogen Energy*. **45**(53), 29546-29559 (2020)

[24] Guo, Q.-L., Li, C., Tu, G.-H., Chen, J.-Q., Wan, B.-B., Liu, Y.: Investigation on the depth effects of the micro-grooves on the suppression of the second modes in the hypersonic boundary layer. *Journal of Physics: Conference Series*. **1786**, 012053 (2021)

[25] Sun, D., Guo, Q.-L., Li, C., Liu, P.-X.: Direct numerical simulation of effects of a micro-ramp on a hypersonic shock wave/boundary layer interaction. *Physics of Fluids*. **31**, 126101 (2019)

[26] Sun, D., Chen, J.-Q., Li, C., Liu, P.-X., Guo, Q.-L., Yuan, X.-X.: On the wake structure of a micro-ramp vortex generator in hypersonic flow. *Physics of Fluids*. **32**, 126111 (2020)

[27] Li, C., Guo, Q.-L., Sun, D., Zhang, H.-X.: Aerothermal Prediction of Hypersonic Flow Around Spherical Capsule Model Using IDDES Approach. *Internationl Journal of Modern Physics B*. **34**, 20400780 (2020)

[28] Wu, M., Martin, M., P.: Direct Numerical Simulation of Supersonic Turbulent Boundary Layer over a Compression Ramp. *AIAA Journal*. **45**(4), 879-889 (2007)

[29] Shur, M., L., Spalart, P., R., Strelets, K., Travin, A.: A hybrid RANS-LES approach with delayed-DES and wall-modelled LES capabilities. *International Journal of Heat and Fluid Flow*. **29**(6), 1638-1649(2008)

[30] Pirozzoli, S., Bernardini, M.: Turbulence in supersonic boundary layers at moderate Reynolds number. *Journal of Fluid Mechanics*. **668**, 120-168 (2011)

[31] Simens, M., P., Jimenez, J., Hoyas, S., Mizuno, Y.: A high-resolution code for a turbulent boundary layers. *Journal of Computational Physics*. **228**, 4218-4231 (2009)

[32] Hutchins, N., Hambleton, W., T., Marusic, I.: Inclined cross-stream stero particle image velocimetry measurements in turbulent boundary layers. *Journal of Fluid Mechanics*. **541**, 21-54 (2005)

[33] Monty, J., P., Stewart, J., A., Williams, R., C., Chong, M., S.: Large-scale features in turbulent pipe and channel flows. *Journal of Fluid Mechanics*. **589**, 147-156 (2007)

[34] Modesti, D., Pirozzoli, S.: Reynolds and Mach number effects in compressible turbulent channel flow. *International Journal of Heat and Fluid Flow*. **59**, 33-49 (2016)

[35] M. Mossi, P. Sagaut. Numerical investigation of fully developed channel flow using shock-capturing schemes. *Computers & Fluids*. 2003, 32: 249-274.

[36] Yaomin Zhao, Zhenhua Xia, Yipeng Shi, Zuoli Xiao, Shiyi Chen. Constrained large-eddy simulation of laminar-turbulent transition in channel flow. *Physics of Fluids*. 2014, 26: 095103.